\documentclass{andromedaone}  

\usepackage[utf8]{inputenc}
\usepackage{amsmath}
\usepackage{amssymb}
\usepackage{amsbsy}
\usepackage{slashed}
\usepackage{xcolor}
\usepackage{graphicx}
\usepackage{url}
\usepackage{cancel}
\usepackage[colorlinks=true,allcolors=darkpurple,pdfborder={0 0 0},linktocpage=false,pdfencoding=auto]{hyperref}
\usepackage{tabularx,booktabs}
\usepackage{multicol}
\usepackage{units}
\usepackage{xspace}
\usepackage[labelfont=bf]{caption}
\usepackage[section]{placeins}
\usepackage{subcaption}
\usepackage{soul} 

\journal{BSM}
\vol{2021}
\jyear{Egypt}
\pages{Zewail} 

\received{xx January 2018}
\published{xx March 2018}

\def\be{\begin{equation}}
\def\ee{\end{equation}}
\def\bea{\begin{eqnarray}}
\def\eea{\end{eqnarray}}

\def\hc{\text{h.c.}}

\def\BR{\text{BR}}
\def\IM{\, \text{Im}}
\def\RE{\, \text{Re}}

\begin{document}

\title{Phenomenology of ultralight scalars in leptonic observables}

\author{Pablo Escribano~\auno{1}}
\address{$^1$Instituto de F\'{i}sica Corpuscular, CSIC-Universitat de Val\`{e}ncia, 46980 Paterna, Spain}

\begin{abstract}
  Ultralight scalars, which are states that are either exactly massless or much lighter than any other massive
  particle in the model, appear in many new physics scenarios. Axions and majorons constitute well-motivated
  examples of this type of particle. In this work, we explore the
  phenomenology of these states in low-energy leptonic
  observables adopting a model independent approach that
  includes both scalar and pseudoscalar interactions. Then, we consider processes in which the ultralight scalar $\phi$ is directly produced, such as $\mu \to e \, \phi$, or acts as a
  mediator, as in $\tau \to \mu \mu \mu$. Finally, contributions to the charged
  leptons magnetic and electric moments are studied
  as well. In particular, it is shown that the muon $g-2$ anomaly can be explained provided a mechanism for suppressing the experimental bounds on the coupling between the ultralight scalar and a pair of muons is introduced.
\end{abstract}

\maketitle


\section{Introduction}

Lepton flavor physics have an exciting experimental perspective in the coming years. Several state-of-the-art experiments recently started taking data and a few more are about to begin~\cite{cs18}. These experiments focus on the search for lepton flavor violating (LFV) processes, which are forbidden in the Standard Model (SM) but strongly motivated by the observation of neutrino flavor oscillations, as well as more precise measurements of lepton flavor conserving observables, such as charged lepton anomalous magnetic moments (AMMs). Specially interesting is the case of the muon observables. The second phase of the MEG experiment, MEG-II~\cite{meg, papa}, is going to guide the search for the radiative LFV decay $\mu \rightarrow e \gamma$, whereas the Mu3e experiment will try to observe the 3-body decay $\mu \rightarrow e e e$ with a sensitivity as low as $10^{-16}$~\cite{papa, mu3e} for the branching ratio. Other searches include neutrinoless $\mu-e$ conversion in nuclei and flavor factories and experiments are also aiming at a large spectrum of flavor observables. Finally, on the flavor conserving side, very recently, the Muon $g - 2$ experiment at Fermilab has presented its long-awaited first results~\cite{g2muFermi}, in agreement with the previous result obtained by the E821 experiment at Brookhaven~\cite{g2muBrook} and hence confirming the long-standing experimental anomaly.

Considering the present hints for the existence of new physics alongside the plethora of promising experiments in the near future, the community of theorists is wondering what kind of new physics can be probed. In this work we study the impact on leptonic observables of ultralight scalars $\phi$ coupling to charged leptons. We will use the term \emph{ultralight scalar} to refer to any generic scalar $\phi$ which is much lighter than the electron, being able to be considered in practice as approximately massless. We will consider a model independent approach by means of effective operators and neglect in all the analytical calculations the mass of the scalar, $m_\phi$. However, notice that this is not an approximation in the case of a Goldstone boson, which is exactly massless. This is the case for the axion and the majoron, two of the most popular ultralight scalars.

\section{Effective Lagrangian}

Although many of the examples of ultralight scalars that can be found in the literature are pseudoscalar particles, the ultralight scalar $\phi$ can have pure scalar couplings as well. This has not been the case in many works in the literature. Therefore, motivated by the seek of generality, we consider here a generic scenario where the ultralight scalar can have both the scalar and pseudoscalar interactions. So, since we are interested in low energy charged leptons processes in the presence of the real scalar, we can generally parametrize the interaction of the ultralight scalar $\phi$ with a pair of charged leptons $\ell_\alpha$ and $\ell_\beta$, with $\alpha, \beta = e, \mu, \tau$, by the effective Lagrangian
\begin{equation}
  \mathcal{L}_{\ell \ell \phi} = \phi \, \overline{\ell}_\beta \left( S_L^{\beta \alpha} P_L + S_R^{\beta \alpha} P_R \right) \ell_\alpha + \hc \, . \label{eq:lagS}
\end{equation}
Here $P_{L,R} = \frac{1}{2} (1 \mp \gamma_5)$ are the usual chiral projectors and no sum over the $\alpha$ and $\beta$ indices is performed. Also, the couplings $S_L$ and $S_R$ are dimensionless and we are taking into account all the possible combinations: $\beta \alpha = \left\{ ee, \mu\mu,\tau\tau,e\mu,e\tau, \mu\tau\right\}$. Finally, note again that although we are considering $\phi$ to be exactly massless for practical reasons, all our results are valid even for massive scalars for which $m_\phi \ll m_e$.

Nevertheless, some LFV observables get contributions from other operators, namely the dipole and 4-fermion operators, which can be found in~\cite{porod14}. Then, the full effective Lagrangian that we will be using is the combination
\begin{equation} \label{eq:lag}
  \mathcal{L} = \mathcal{L}_{\ell \ell \phi} + \mathcal{L}_{\ell \ell \gamma} + \mathcal{L}_{4 \ell} \, ,
\end{equation}
with
\begin{align}
  \mathcal{L}_{\ell \ell \gamma} & = \frac{e \, m_\alpha}{2} \, \overline{\ell}_{\beta} \, \sigma^{\mu \nu} \left[ \left( K_2^L \right)^{\beta \alpha} P_L + \left( K_2^R \right)^{\beta \alpha} P_R \right] \ell_{\alpha} F_{\mu \nu} + \hc \, , \label{eq:lagA} \\
  \mathcal{L}_{4 \ell} & = \sum_{I=S, V, T \atop X, Y=L, R} \left( A_{X Y}^{I} \right)^{\beta \alpha \delta \gamma} \, \overline{\ell}_\beta \Gamma_I P_X \ell_{\alpha} \, \overline{\ell}_\delta \Gamma_I P_Y \ell_\gamma + \hc \, , \label{eq:lag4F}  
\end{align}
where $F_{\mu \nu} = \partial_\mu A_\nu - \partial_\nu A_\mu$ is the usual electromagnetic field strength tensor, with $A_\mu$ the photon field, and we have defined the tensors $\Gamma_S = 1$, $\Gamma_V = \gamma_{\mu}$ and $\Gamma_T = \sigma_{\mu \nu}$. Again, we do not sum over the charged lepton flavor indices in Eqs.~\eqref{eq:lagA} and \eqref{eq:lag4F}, and all the new couplings have dimensions of mass$^{-2}$. Finally, we are normalizing the Lagrangian in Eq.~\eqref{eq:lagA} by including the mass of the heaviest charged lepton in the process of interest.

In the following, we will concentrate on purely leptonic observables, disregarding interactions between the ultralight scalar and quarks. Using the analytical expressions already shown in~\cite{ultral}, we will discuss some phenomenological aspects of several leptonic observables, such as the decays $\ell_\alpha \to \ell_\beta \, \phi$ or $\ell_\alpha \to \ell_\beta \ell_\beta \ell_\beta$, or the electron and muon anomalous electric and magnetic dipole moments.

\section{Phenomenological discussion}

As mentioned above, all the complete analytical expressions have been obtained in~\cite{ultral}. Here we will only show the approximated expressions and we will restrict ourselves to the phenomenological implications of the observables computed.

\subsection[Searches for \texorpdfstring{$\ell_\alpha \to \ell_\beta \, \phi$}{\unichar{"2113}\unichar{"03B1} \unichar{"2192} \unichar{"2113}\unichar{"03B2} \unichar{"03D5}}]{Searches for $\boldsymbol{\ell_\alpha \to \ell_\beta \, \phi}$}

The importance of this observable lies in the fact that it can be used to obtain the most constraining experimental bounds on the flavor violating $S_A^{\beta \alpha}$ couplings. Also, the very simple expression for the decay width,
\begin{equation}
  \Gamma \left(\ell_\alpha \to \ell_\beta \, \phi \right) = \frac{m_\alpha}{32 \, \pi} \left( \left| S_L^{\beta \alpha} \right|^2 + \left| S_R^{\beta \alpha} \right|^2 \right) \, ,
  \label{eq:decaywidth_betaphi}
\end{equation}
allows us to make straightforward derivations. Note that terms proportional to the small ratio $m_\beta / m_\alpha$ have been neglected in the expression.

Starting first with muon decays, the current strongest bound on the branching ratio for the process $\mu^+ \rightarrow e^+ \phi$ was obtained at TRIUMF with a muon beam highly polarized in the direction opposite to the muon momentum and concentrating the search in the forward region, as explained in~\cite{hirsch09}. Therefore, the limit found was $\BR \left(\mu \to e \, \phi\right) < 2.6 \times 10^{-6}$ at 90\% C.L.~\cite{Jodidio:1986mz}, and it was valid only when $S_L^{e \mu} = 0$. However, the authors of~\cite{hirsch09} obtained the conservative bound $\BR \left(\mu \to e \, \phi\right) \lesssim 10^{-5}$, which is valid for any chirality of the couplings, using the data shown in~\cite{Jodidio:1986mz}. More recently, a similar bound was obtained by the TWIST collaboration~\cite{twist}. So, using this result, one finds the upper limit 
\begin{equation} \label{eq:limemu1}
  \left| S^{e \mu} \right| < 5.3 \times 10^{-11} 
\end{equation}
on the couplings, where we have defined the convenient combination
\begin{equation}
  \left| S^{\beta \alpha} \right| = \left( \left| S^{\beta \alpha}_L \right|^2 + \left| S^{\beta \alpha}_R \right|^2 \right)^{1/2} \, .
\end{equation}
Regarding to $\tau$ decays, the ARGUS collaboration found
\begin{equation}
  \frac{\BR\left(\tau \to e \, \phi \, \right)}{\BR \left(\tau \to e \, \nu \, \bar{\nu} \right) } < 0.015 \quad , \quad  \frac{\BR\left(\tau \to \mu \, \phi \right)}{\BR \left(\tau \to \mu \, \nu \, \bar{\nu} \right) } < 0.026 \, ,
\end{equation}
at 95\% C.L.. Although these bounds are the actual best limits, they are milder than the ones for muons. But they still allow us to derive constricting bounds on the $\tau$ couplings:
\begin{equation} \label{eq:taulim}
  \begin{split}
    & \left| S^{e \tau} \right| < 5.9 \times 10^{-7} \, , \\
    & \left| S^{\mu \tau} \right| < 7.6 \times 10^{-7} \, .
  \end{split}
\end{equation}
Finally, these limits will likely be improved at Belle II.

\subsection[\texorpdfstring{$\ell_\alpha \to \ell_\beta \, \gamma \, \phi$}{\unichar{"2113}\unichar{"03B1} \unichar{"2192} \unichar{"2113}\unichar{"03B2} \unichar{"03B3} \unichar{"03D5}} at the MEG experiment]{$\boldsymbol{\ell_\alpha \to \ell_\beta \, \gamma \, \phi}$ at the MEG experiment}

\begin{figure}
\centering\includegraphics[scale=0.45]{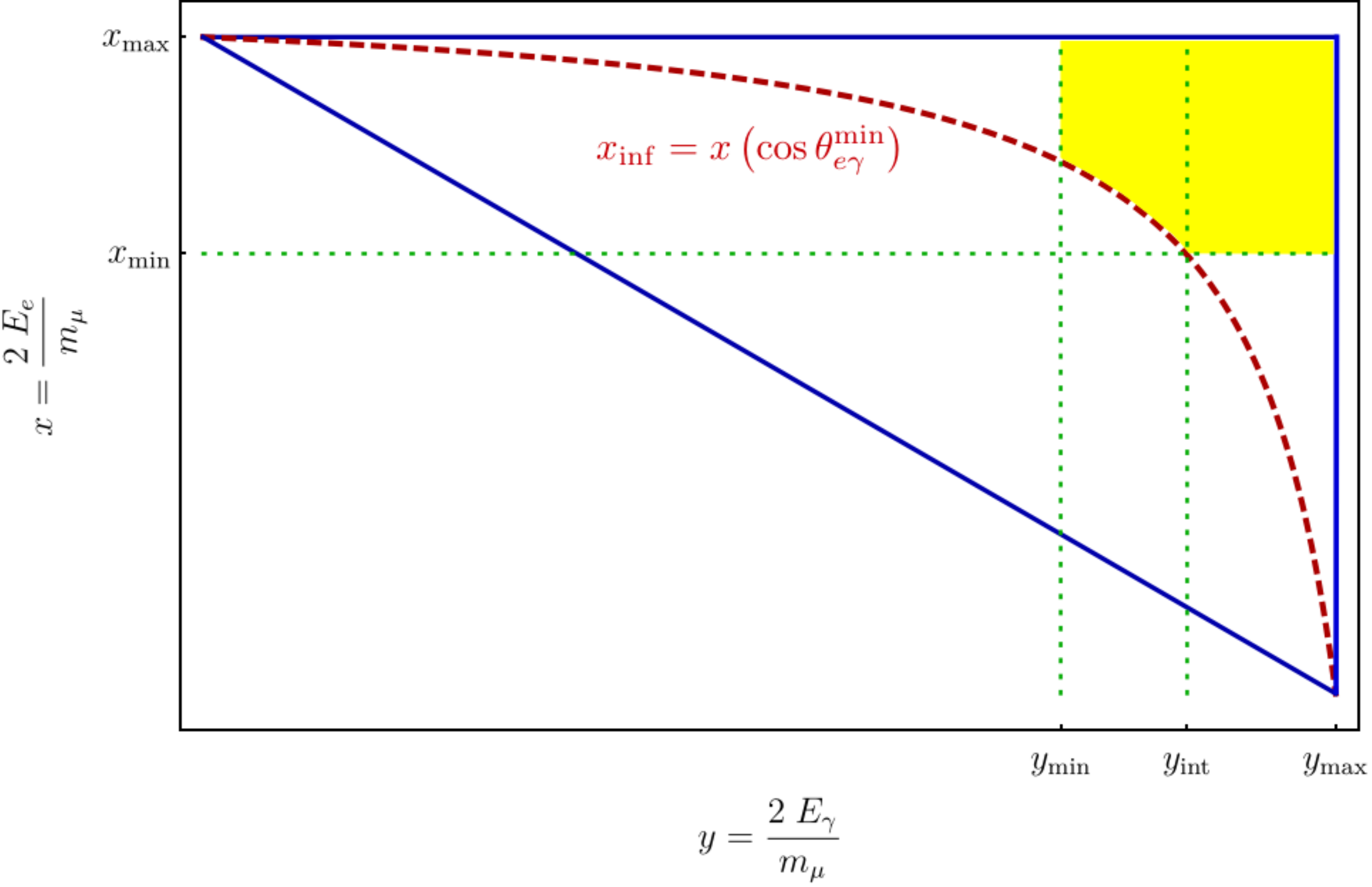}
\caption{Not realistic illustration of the allowed phase space region for the
    process $\mu \to e \, \gamma \, \phi$ due to a given experiment. 
    The total phase space in principle available by kinematics is represented by the blue continuous lines, which are given by $\cos \theta_{e \gamma} = \pm 1$ in Eq.~\eqref{eq:theta}. The red dashed curve
    represents $x_{\rm inf}(y)$, which corresponds to the minimal
    $\theta_{e \gamma}$ angle given the experiment, and it excludes
    the region below it. The green dotted straight lines at $x_{\min}$
    and $y_{\min}$ are the minimal positron and photon energy,
    respectively, that the experiment can measure, while $y_{\rm int}$
    is the value of $y$ for which $x_{\min}$ and $x_{\rm inf}$
    intersect. Finally, the yellow surface is the region over which we must
    integrate.
    \label{fig:phase-space}}
\end{figure}

Additional bounds to the $e - \mu$ couplings can be set using the observable $\mu \to e \, \gamma \, \phi$ alongside the results of the MEG experiment. The decay width of the process can be written as
\begin{equation}
  \Gamma \left( \ell_\alpha \to \ell_\beta \, \gamma \, \phi \right) = \frac{\alpha \, m_\alpha}{64 \pi^2} \left( \left| S^{\beta \alpha}_L \right|^2 + \left| S^{\beta \alpha}_R \right|^2 \right) \mathcal{I} \left( x_{ \min } , y_{ \min } \right) \,.
  \label{eq:decaywidth_betagammaphi}
\end{equation}
Here, terms proportional to $m_\beta / m_\alpha$ have been neglected and $\mathcal{I} \left( x_{ \min } , y_{ \min } \right)$ is a phase space integral given by
\begin{equation}
  \mathcal{I} \left( x_{ \min } , y_{ \min } \right) = \int \text{d} x \, \text{d} y \, \frac{ \left( x - 1 \right) \left( 2 - x y - y \right) }{ y^{2} \left( 1 - x - y \right) } \, ,
  \label{eq:phase_space_integral}
\end{equation}
where we have defined the useful dimensionless parameters $x$ and $y$, given by
\begin{equation}
  x = \frac{2 E_{\beta}}{m_{\alpha}} \quad, \quad y = \frac{2 E_{\gamma}}{m_{\alpha}} \, .
\end{equation}
Notice that, together with $z = 2 E_{\phi} / m_{\alpha}$, they must satisfy condition $x + y + z = 2$.

\begin{figure}
  \begin{subfigure}{0.5\textwidth}
    \centering
    \includegraphics[scale=0.25]{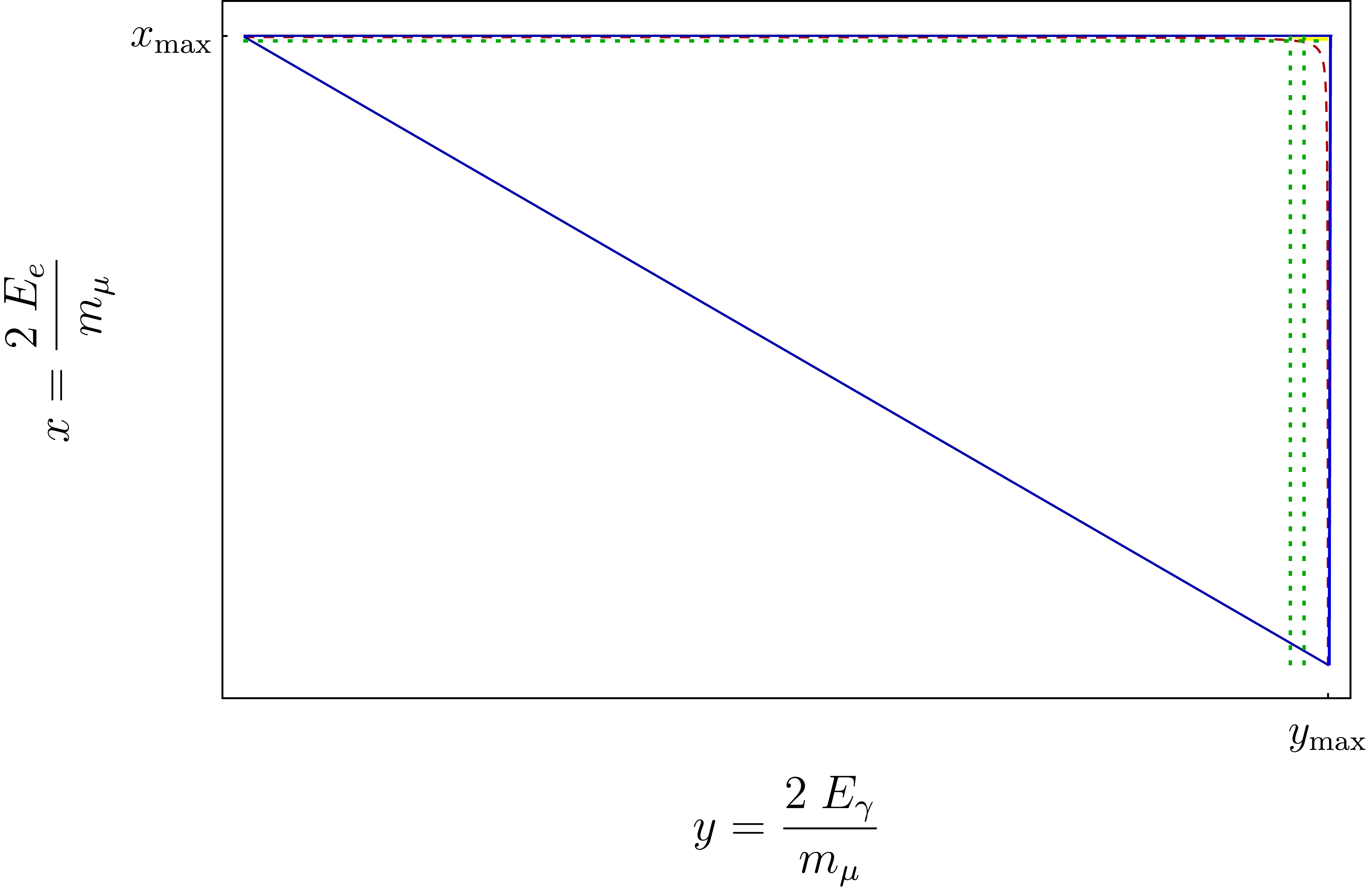}
  \end{subfigure}
  \begin{subfigure}{0.5\textwidth}
    \centering
    \includegraphics[scale=0.25]{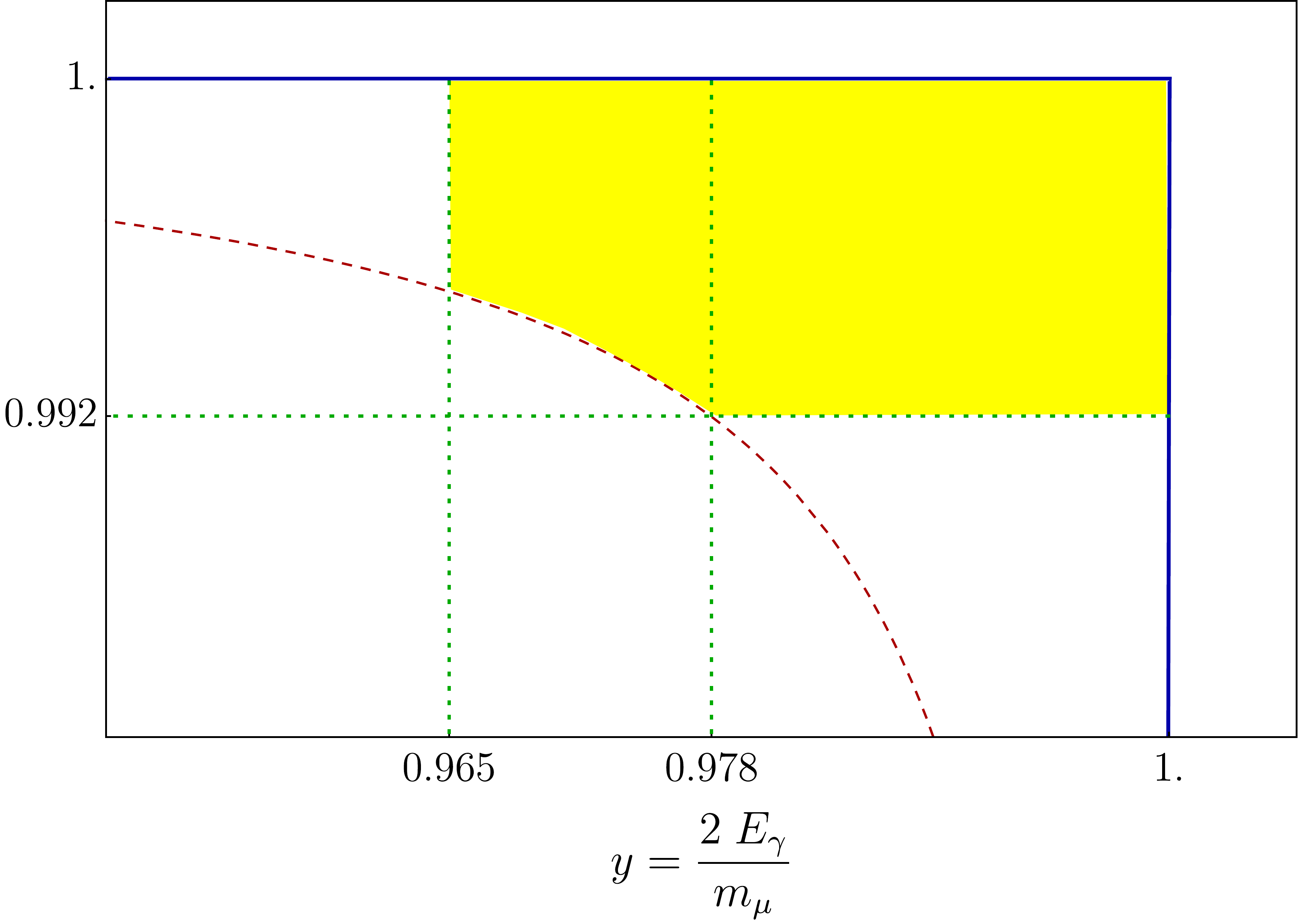}
  \end{subfigure}
  \caption{Realistic version of the phase space region limited by the
    MEG experimental cuts, given in
    Eq.~\eqref{eq:MEGcuts}. A zoom of the figure on the left, centered on the yellow area is shown on the figure on the right.}
  \label{fig:phase-space-realistic}
\end{figure}

On the other hand, since the MEG experiment was designed specifically for the $\mu \rightarrow e \gamma$ decay, it concentrates on $E_e \simeq m_\mu / 2$ and $\cos \theta_{e \gamma} \simeq -1$ corresponding to the positron and the photon emitted back to back. However, and luckily for our aim here, the experimental resolution is finite, making MEG sensitive to our decay of interest. Thus, the experimental cuts with which the final MEG results were obtained are~\cite{Mori2016vwi}
\begin{equation} \label{eq:MEGcuts}
  \cos \theta_{e \gamma} < - 0.99963 \quad, \quad 51.0 < E_\gamma < 55.5 \, \text{MeV} \quad, \quad 52.4 < E_e < 55.0 \, \text{MeV} \, .
\end{equation}
These bounds serve to define the kinematical region for the calculation of the previous phase space integral. The idea is that events of the 3-body decay that fall into this region will be detected in the experiment. Now, it proves convenient to divide the kinematical region into two subregions,
\begin{equation}
  \begin{split}
    y_{\min} = \frac{2 \, E_\gamma^{\min}}{m_\mu} < y  < y_{\rm int} \quad  , \quad x_{\rm inf} < x & < x_{\max} = 1 \, ,
  \end{split}
\end{equation}
and
\begin{equation}
  \begin{split}
    y_{\rm int} < y  < y_{\max} = 1  \quad  , \quad x_{\min} = \frac{2 \, E_e^{\min}}{m_\mu} < x & < x_{\max} \, ,
  \end{split}
\end{equation}
where $x_{\rm inf} = x_{\rm inf}(y)$ is the value of $x$ such that
$\cos \theta_{e \gamma} = \cos \theta_{e \gamma}^{\min}$ for each
value of $y$, where $\theta_[e \gamma]^{\min}$ is given by the upper bound on the cosine in~\eqref{eq:MEGcuts}. This can be easily found by solving
\begin{equation} \label{eq:theta}
\cos \theta_{e \gamma} = 1 + \frac{2 - 2(x+y)}{xy} \, .
\end{equation}
Finally $y_{\rm int}$ is the value of $y$ for which $x_{\min}$ and $x_{\rm inf}$ coincide. These subregions are shown in Fig.~\ref{fig:phase-space} with the experimental restrictions modified, enlarging the kinematical region available for the sake of clarity. However, a realistic representation obtained with the cuts in Eq.~\eqref{eq:MEGcuts} is shown in Fig.~\ref{fig:phase-space-realistic}. In this last figure, it is clearly seen that we have a strong suppression due to the phase space integral, which can be numerically computed,
\begin{equation}
  \mathcal{I} \left( x_{ \min } , y_{ \min } \right)_{\rm MEG} = 3.8 \times 10^{-8} \, .
  \label{eq:phase_space_integra_num}
\end{equation}
Finally, plugging this result into Eq.~\eqref{eq:decaywidth_betagammaphi}, we have that the branching ratio restricted to the MEG experimental restrictions is
\begin{equation}
  \text{BR}_{\rm MEG} \left( \mu \to e \, \gamma \, \phi \right) = 1.5 \times 10^{5}  \left( \left| S^{e \mu}_L \right|^2 + \left| S^{e \mu}_R \right|^2 \right) \, ,
  \label{eq:num_BRegammamu}
\end{equation}
and combining with the MEG results, which require $\BR \left( \mu \to e \, \gamma \right) < 4.2
\times 10^{-13}$~\cite{Mori2016vwi}, leads to
\begin{equation} \label{eq:limemu2}
	\left| S^{e \mu} \right| < 1.6 \times 10^{-9} \, ,
\end{equation}
since the the MEG bound must also be satisfied by $\text{BR}_{\rm MEG} \left( \mu \to e \, \gamma \, \phi
\right)$. Notice that this bound is much worse than the one given in Eq.~\eqref{eq:limemu1}. This, however, was expected due to the strong phase space suppression at MEG, which was not designed to search for $\mu \to e \, \gamma \, \phi$. Nevertheless, as shown in~\cite{ultral}, one can obtain more stringent bounds using the results of the Crystal Box experiment at LAMPF~\cite{Bolton1986tv,Goldman1987hy,Bolton1988af}, that lead to 
\begin{equation}
  \left| S^{e \mu} \right| < 9.5 \times 10^{-11} \,  .
\end{equation}
But it is still less stringent than the bound obtained in the previous section.

\subsection[\texorpdfstring{$\ell_\alpha \to \ell_\beta \, \gamma$}{\unichar{"2113}\unichar{"03B1} \unichar{"2192} \unichar{"2113}\unichar{"03B2} \unichar{"03B3}} vs \texorpdfstring{$\ell_\alpha \to \ell_\beta \ell_\beta \ell_\beta$}{\unichar{"2113}\unichar{"03B1} \unichar{"2192} \unichar{"2113}\unichar{"03B2} \unichar{"2113}\unichar{"03B2} \unichar{"2113}\unichar{"03B2}}]{$\boldsymbol{\ell_\alpha \to \ell_\beta \gamma}$ vs $\boldsymbol{\ell_\alpha \to \ell_\beta \ell_\beta \ell_\beta}$}

Now we will focus on two LFV decays at the same time, $\ell_\alpha \to \ell_\beta \, \gamma$ and $\ell_\alpha \to \ell_\beta \ell_\beta \ell_\beta$, since they can be used, if observed, in a complementary way to probe what kind of new physics we have. The first process can only be induced by dipole operators, having its decay width a very simple form,
\begin{equation} \label{eq:widthLLG}
  \Gamma \left( \ell_\alpha \to \ell_\beta \gamma \right) = \frac{e^2 \, m_\alpha^5}{16 \, \pi} \left[ \left|\left(K_2^L\right)^{\beta\alpha}\right|^2 + \left|\left(K_2^R\right)^{\beta \alpha}\right|^2 \right] \, .
\end{equation}
However, the second decay receives contributions from both dipole and non-dipole operators, having a very complicated expression. 

\subsubsection*{General dipole contributions}

First, we contemplate the general scenario with dipole contributions independent of the non-dipole ones that can be induced by the ultralight scalar. Actually, this happens when there exist sources of LFV not related to $\phi$. For the discussion here, we will drop all the 4-fermion operators from the Lagrangian in Eq.~\eqref{eq:lag} as well as all the right-handed  photonic dipole and scalar-mediated operators. Therefore, any contribution to the observables can only come from this effective Lagrangian:
\begin{equation} \label{eq:lagSimp}
  \mathcal{L}_{\rm LFV}^{\rm simp} = \frac{e \, m_\alpha \, \left(K_2^L\right)^{\beta \alpha}}{2} \, \overline{\ell}_{\beta} \, \sigma^{\mu \nu} \, P_L \, \ell_{\alpha} F_{\mu \nu} + S_L^{\beta \alpha} \, \phi \, \overline{\ell}_\beta \, P_L \, \ell_\alpha + \hc \, .
\end{equation}
In the following, and until the next section, we will assume that $S_L^{\beta \beta} = S_L^{\beta \beta}$. Hence, in this simplified scenario the decay width of the decay to three charged leptons takes the simple form
\begin{equation}
  \begin{split}
    &\Gamma \left(\ell_{\alpha}^{-} \rightarrow \ell_{\beta}^{-} \ell_{\beta}^{-} \ell_{\beta}^{+}\right)=  \frac{m_{\alpha}}{512 \pi^{3}} \Biggl\{ \left|S_{L}^{\beta \alpha}\right|^{2} \left\{\left|S_L^{\beta \beta}\right|^{2}\left(4 \log \frac{m_{\alpha}}{m_{\beta}}-\frac{49}{6}\right)-\frac{2}{6}\left[\left(S_L^{\beta \beta *}\right)^{2}+\left(S_L^{\beta \beta}\right)^{2}\right]\right\}\Biggr. \Biggl. + \, m_\alpha^4 \, e^{4} \left|K_{2}^{L}\right|^{2} \left(\frac{16}{3} \log \frac{m_{\ell_{\alpha}}}{m_{\ell_{\beta}}}-\frac{22}{3}\right) \Biggr\} \, .
  \end{split}
\end{equation}
Now, we make use of a useful parametrization of our remaining couplings. So, inspired by~\cite{gouvea2013}, we define
\begin{equation} \label{eq:param}
  e \, \left(K_2^L\right)^{\beta \alpha} \equiv \frac{1}{\left(\kappa + 1 \right) \Lambda^2} \, , \qquad  S_L^{\beta \alpha} \equiv m_\alpha \, \frac{\kappa}{\left(\kappa + 1 \right) \Lambda} \, ,
\end{equation}
where $\Lambda$, which has dimensions of mass, is meant to represent the effective mass scale at which the coefficients $K_2^2$ and $S_L$ are induced. Besides, $\kappa$ is a dimensionless parameter that accounts for the relative size of both operators in~\eqref{eq:lagSimp}. If $\kappa \gg 1$, the scalar-mediated contribution is dominant, while in the opposite case, the photonic dipole contribution is more important. Finally, note that we are normalizing $S_L$ by including the mass of the heaviest charged lepton in the process we are calculating. Nevertheless, this is only done in this particular analysis. In the rest of this work there is no hierarchy assumed among the $S_{L,R}^{\beta \alpha}$ couplings.

\begin{figure}
\centering\includegraphics[scale=0.35]{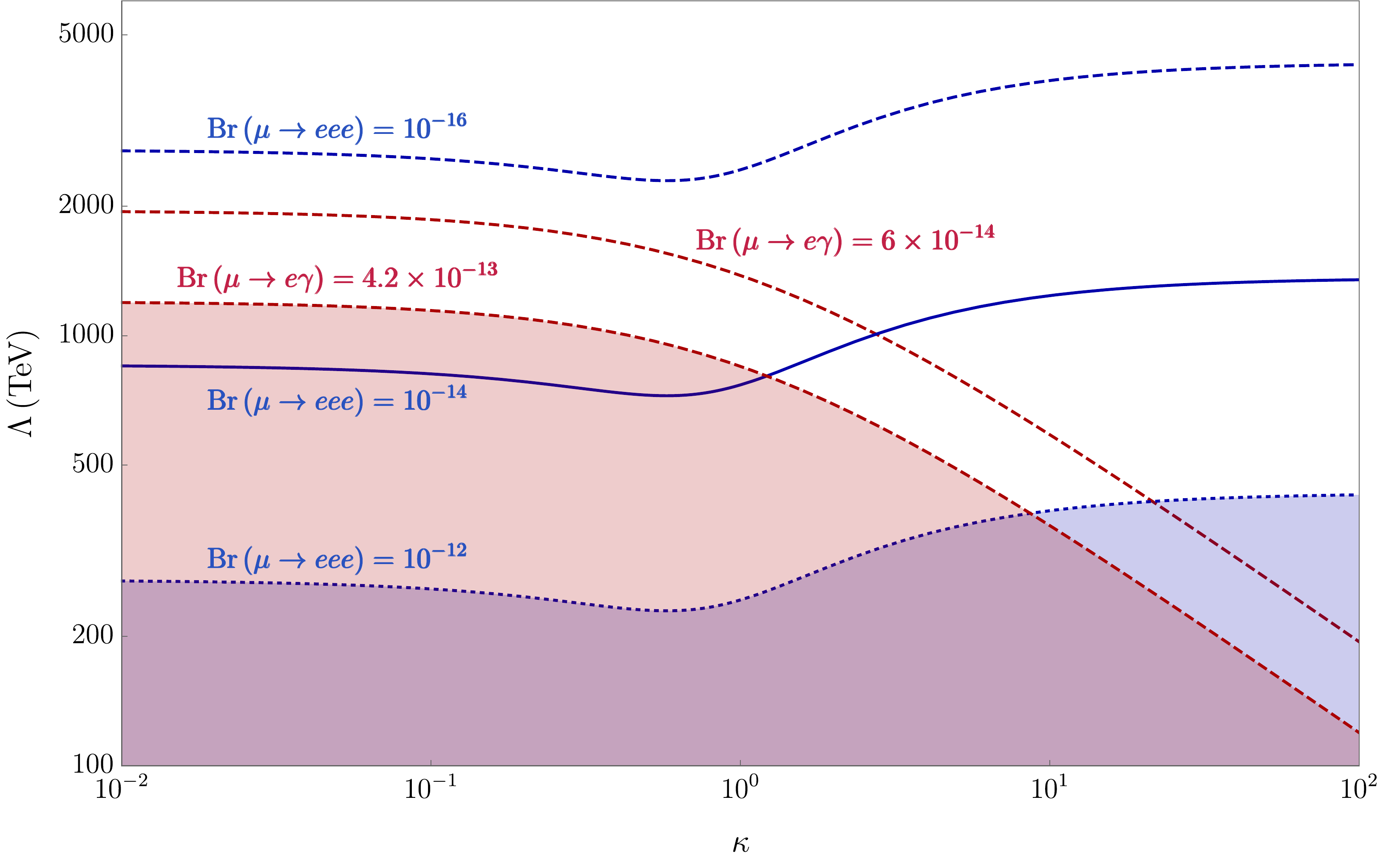}
\caption{Contours of $\BR(\mu \to e \gamma)$ and $\BR(\mu \to eee)$
    in the $\kappa$-$\Lambda$ parameter space using Eq.~\eqref{eq:decaysnewparam}. The lowest values in both red and blue are the future sensitivities for the MEG-II and Mu3e experiments, respectively,
    while the current bounds $\BR(\mu \to e \gamma) < 4.2 \cdot 10^{-13}$ and $\BR(\mu \to eee)
    < 10^{-12}$~\cite{Tanabashi2018oca} exclude the colored surfaces.
    \label{fig:L3L} }
\end{figure}

\begin{figure}
  \begin{subfigure}{0.5\textwidth}
    \centering
    \includegraphics[scale=0.4]{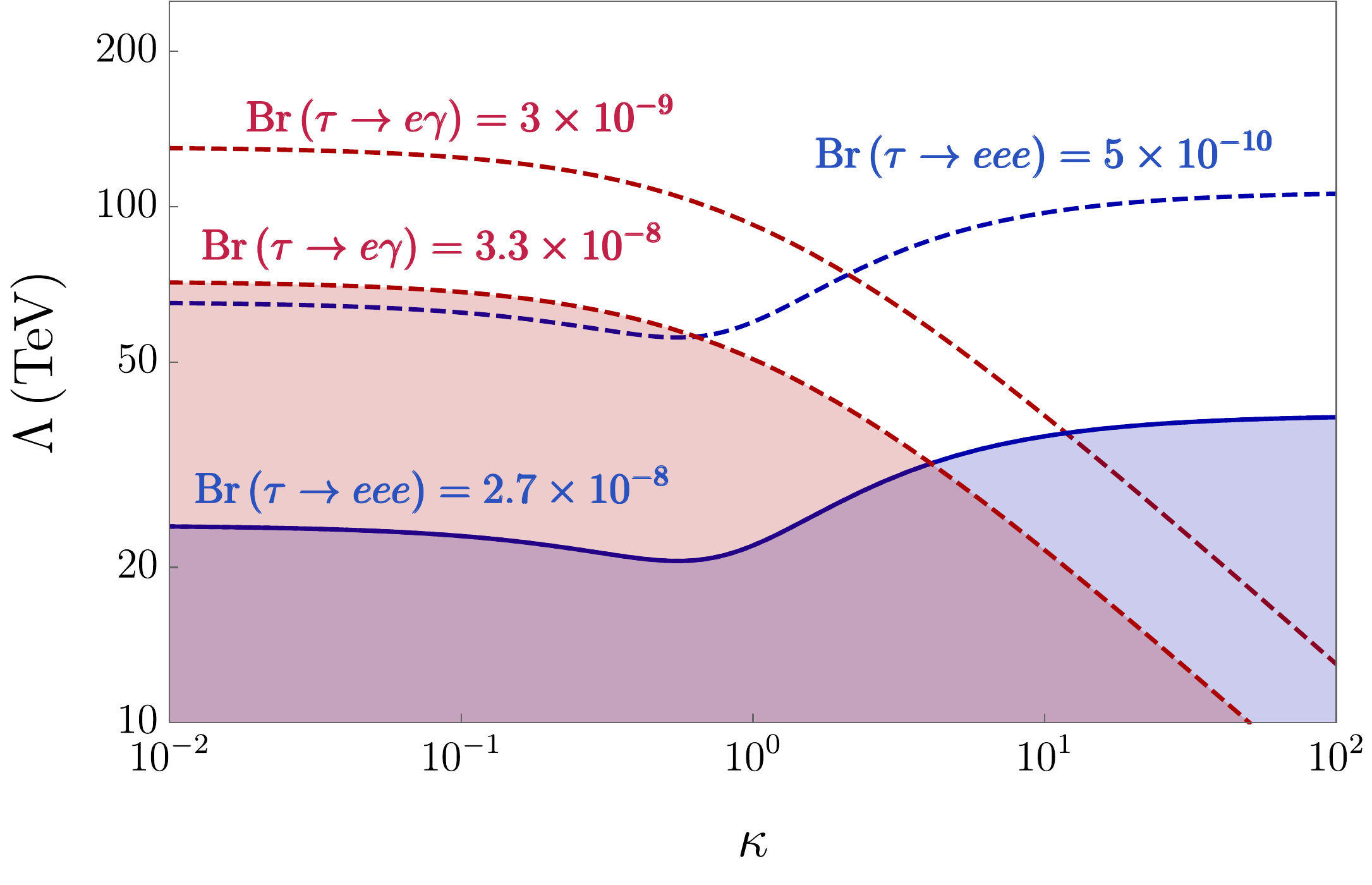}
  \end{subfigure}
  \begin{subfigure}{0.5\textwidth}
    \centering
    \includegraphics[scale=0.4]{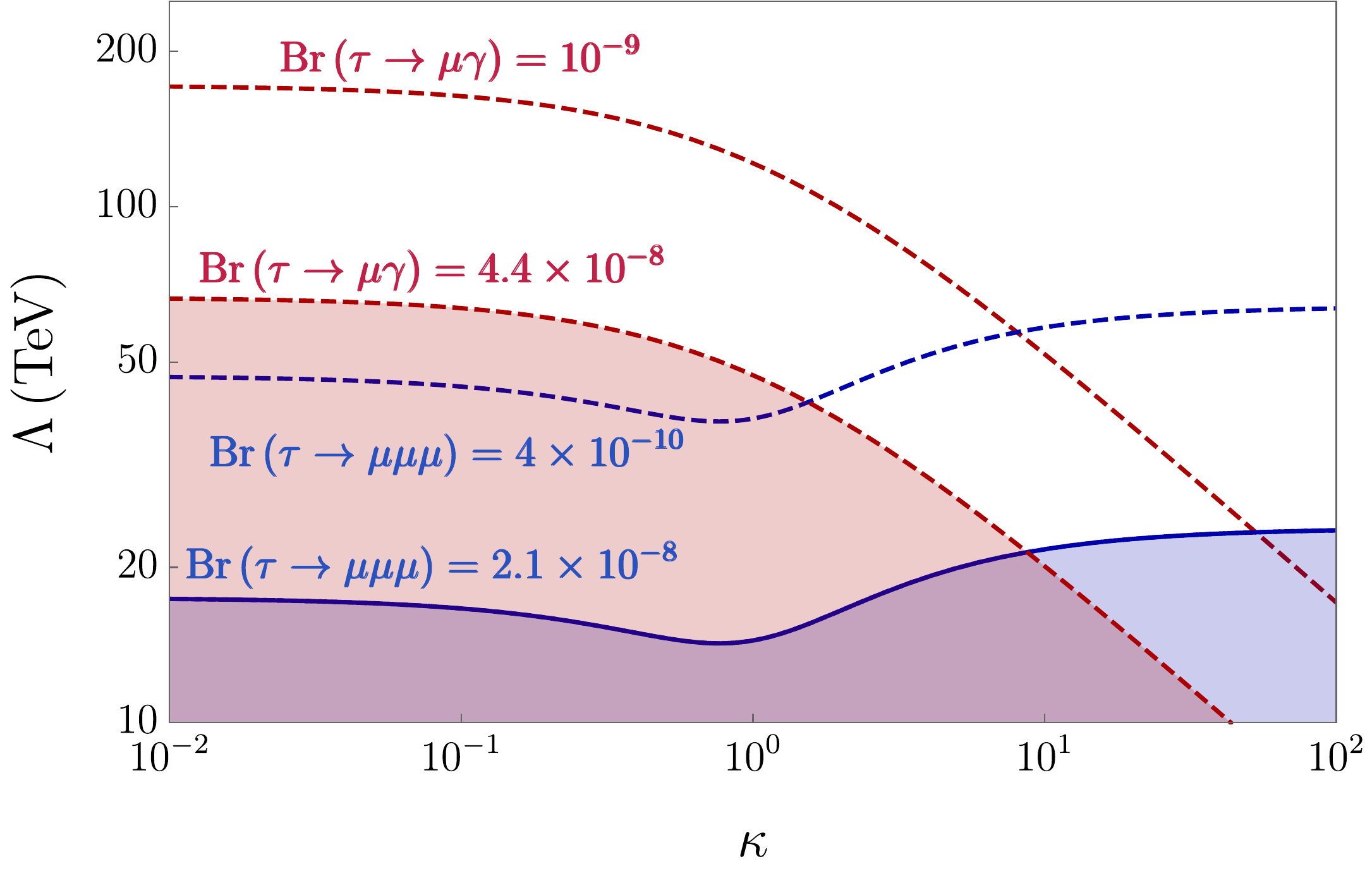}
  \end{subfigure}
  \caption{Contours of $\BR(\tau \to e \gamma)$ and $\BR(\tau \to
    eee)$, on the left, and $\BR(\tau \to \mu \gamma)$ and $\BR(\tau
    \to \mu \mu \mu)$, on the right, in the $\kappa$-$\Lambda$
    parameter space using Eq.~\eqref{eq:decaysnewparam}. The lowest values in both red and blue are the 
    future sensitivities of the Belle II experiment~\cite{Perez2019cdy},
    while the current bounds $\BR(\tau \to e \gamma) < 3.3 \cdot 10^{-8}$, $\BR(\tau \to \mu
    \gamma) < 4.4 \cdot 10^{-8}$, $\BR(\tau \to eee) < 2.7 \cdot
    10^{-8}$ and $\BR(\tau \to \mu \mu \mu) < 2.1 \cdot
    10^{-8}$~\cite{Tanabashi2018oca} exclude the colored surfaces.
  \label{fig:L3L-tau}}
\end{figure}

The expressions of the processes in the new parametrization are written as
\begin{equation}
  \begin{split}
    &\Gamma \left(\ell_{\alpha}^{-} \rightarrow \ell_{\beta}^{-} \ell_{\beta}^{-} \ell_{\beta}^{+}\right)= \frac{m_{\alpha}^5}{512 \pi^{3}} \left[ \frac{\kappa^2}{\left(\kappa + 1 \right)^4 \Lambda^4}\left(4 \log \frac{m_{\alpha}}{m_{\beta}}-\frac{53}{6}\right) + \frac{e^2}{\left( \kappa + 1 \right)^4 \Lambda^4} \left(\frac{16}{3} \log \frac{m_{\ell_{\alpha}}}{m_{\ell_{\beta}}}-\frac{22}{3}\right) \right] \, , \\
    & \Gamma\left(\ell_{\alpha} \rightarrow \ell_{\beta} \gamma\right)=\frac{m_{\alpha}^{5}}{16 \pi} \frac{1}{\left( \kappa + 1 \right)^2 \Lambda^4} \, ,
  \label{eq:decaysnewparam}
  \end{split}
\end{equation}
and with them, we show in Fig.~\ref{fig:L3L} both $\BR(\mu \to e \gamma)$ and $\BR(\mu \to eee)$ as a function of the new parameters. We can extract that if $\BR(\mu \to eee) > 10^{-16}$ is observed and $\kappa \gg 1$, $\Lambda$ would be bounded from above by approximately $3000 \, \text{TeV}$. On the other hand, we find a slightly lower limit if $\BR(\mu \to e \gamma) > 10^{-14}$ when $\kappa \ll 1$. The limits used here are the expected final sensitivities of the MEG-II and Mu3e experiments. In addition, notice that the searches for the 3-body decay in Mu3e will be very constraining for $\Lambda$ in all the range of $\kappa$. Very similar results are shown for $\tau$ decays in Fig.~\ref{fig:L3L-tau}. For these observables it is expected that the experimental limits will be improved by an order of magnitude by the LHCb and Belle II collaborations. Interestingly, the observation of $\tau \rightarrow e e e$ at Belle II is forbidden by the current limit on $\BR(\tau \to e \gamma)$ in the case of $\kappa \ll 1$. Therefore, if Belle II finds the decay, a larger value of $\kappa$ would be necessary. Finally, we have the same qualitative results for $\tau \to \mu$ decays.

\subsubsection*{$\boldsymbol{\phi}$-induced dipole contributions}

\begin{figure}
\centering\includegraphics[scale=0.45]{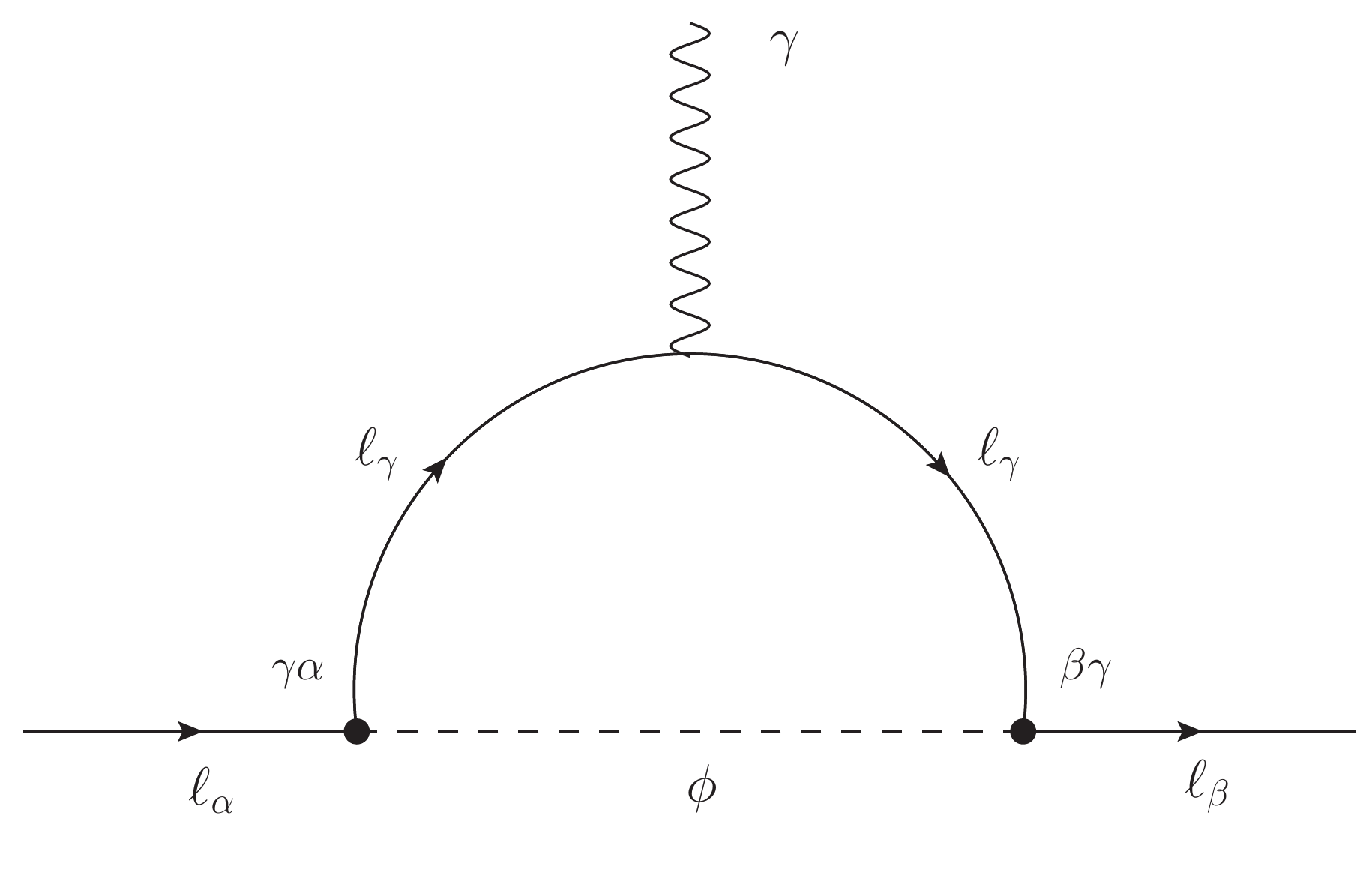}
\caption{One-loop diagram contributing to the process $\ell_{\alpha} \to \ell_{\beta} \gamma$ in the presence of the effective Lagrangian in Eq.~\eqref{eq:lagS}. The greek letters in the vertices represent the flavor indices of the couplings contributing to the diagram.
  \label{fig:Diagrambeta_gamma}}
\end{figure}

In this second part of the section, we are taking a different approach. We are considering that the photonic dipole operators are induced by loops involving the ultralight scalar $\phi$, as it is shown in Fig.~\ref{fig:Diagrambeta_gamma}. So, in this scenario, we assume that the scalar's contribution to dipole operators is the dominant or the only one. To make things simpler, we will only allow the couplings involving electrons, that is $S^{ee}$ and $S_{L,R}^{e \mu}$, to be non-zero (and also real) in the following analysis. 

Under these assumptions and expanding at first order in the mass of the electron, $m_e$, we have these expressions for the dipole couplings:
\begin{align}
  \left(K_2^L\right)^{e \mu} = &  \frac{S^{ee}}{96 \pi^2 \, m_\mu^3} \biggl\{ 3 \, m_\mu \, S_R^{e \mu} + m_e \left( -6 \, S_L^{e \mu} + 2 \, \pi^2 \, S_L^{e \mu} + 3 \, S_R^{e \mu} \right)  + \, 3 \, m_e S_L^{e \mu} \log \left( - \frac{m_e^2}{m_\mu^2} \right) \left[ 1 + \log \left( - \frac{m_e^2}{m_\mu^2} \right) \right] \biggr\} \, , 
\end{align}
\begin{align}
  \left(K_2^R\right)^{e \mu} = & \frac{S^{ee}}{96 \pi^2 \, m_\mu^3} \biggl\{ 3 \, m_\mu \, S_L^{e \mu} + m_e \left( -6 \, S_R^{e \mu} + 2 \, \pi^2 \, S_R^{e \mu} + 3 \, S_L^{e \mu} \right) + \, 3 \, m_e S_R^{e \mu} \log \left( - \frac{m_e^2}{m_\mu^2} \right) \left[ 1 + \log \left( - \frac{m_e^2}{m_\mu^2} \right) \right] \biggr\} \, .
\end{align}
With them, and  defining the useful mass ratio $r = \frac{m_\mu^2}{m_e^2}$, we are able to compute the ratio
\begin{equation} \label{eq:Rmue}
  R_{\alpha \beta} = \frac{\BR(\ell_\alpha \to \ell_\beta \ell_\beta
    \ell_\beta)}{\BR(\ell_\alpha \to \ell_\beta \, \gamma)} \, .
\end{equation}
for some simplified scenarios:
\begin{itemize}

\item Scenario 1: $S_L^{e \mu} = 0$ or $S_R^{e \mu} = 0$ 

\begin{equation} 
  R_{\mu e}^{(1)} \approx \frac{4 \, \pi \, r}{3 \, \alpha} \, \frac{12 \, \log r - 53}{|\log (-r)|^4 + r} \approx 3.2 \cdot 10^4 \, .
\end{equation} 

\item Scenario 2: $S_L^{e \mu} = S_R^{e \mu}$

\begin{equation}
  R_{\mu e}^{(2)} \approx \frac{4 \, \pi \, r}{3 \, \alpha} \, \frac{12 \, \log r - 53}{|\log^2 (-r) + \sqrt{r}|} \approx 1.9 \cdot 10^4 \, .
\end{equation} 

\item Scenario 3: $S_L^{e \mu} = - S_R^{e \mu}$

\begin{equation}
  R_{\mu e}^{(3)} \approx \frac{4 \, \pi \, r}{3 \, \alpha} \, \frac{12 \, \log r - 53}{|\log^2 (-r) - \sqrt{r}|} \approx 1.1 \cdot 10^5 \, .
\end{equation} 

\end{itemize}
From the obtained values, we find that $R_{\mu e} \gg 1$ in all the cases, as expected. The decay $\ell_\alpha \to \ell_\beta \ell_\beta \ell_\beta$ is induced at tree-level by exchange of the ultralight scalar, whereas $\ell_\alpha \to \ell_\beta \, \gamma$ can only take place at loop order. Also, and more interestingly, we obtained different predictions for the ratio depending on the particular scenario. Then, this could allow us to determine the nature of the scalar if both processes are observed.

\subsection{Lepton magnetic and electric dipole moments}

Currently, there is a long-standing discrepancy between the Standard Model prediction for the electron and muon anomalous magnetic moments and their experimental determination. Indeed, due to the recent publication of the results of the Muon $g - 2$ experiment at Fermilab~\cite{g2muFermi}, the deviation in the case of the muon has become more relevant.
\begin{align}
  \Delta a_e &= a_e^{\text{exp}} - a_e^{\text{SM}} = (-87 \pm 36) \times 10^{-14} \, , \\
  \Delta a_{\mu} &= a_\mu^{\text{exp}} - a_\mu^{\text{SM}} = (25.1 \pm 5.9) \times 10^{-10} \, ,
\end{align}
where
\begin{equation}
  a_\beta = \frac{g_\beta - 2}{2} \, .
\end{equation}
In the case of the muon AMM, the anomaly has been updated to the level of $ 4.2 \, \sigma $, while for the electron AMM it is a bit lower, slightly below $\sim 3 \, \sigma$. These deviations from the theoretical predictions can be interpreted as possible hints of new physics, as we will do in this section. Nevertheless, to fully confirm the anomalies, we still require more measurements and, possibly, improved theoretical calculations. Concerning the EDMs, the SM predictions for these observables are well beyond the experimental sensitivities in the near future. Therefore, any positive signal of them in an experiment would be a clear indication of new physics effects, which, in addition, must violate CP. The current best limits for the electron and muon EDMs are~\cite{Bennett2008dy,Andreev2018ayy} 
\begin{align}
  |d_e| &< 1.1 \times 10^{-29} \, e \, \text{cm} \, , \\
  |d_\mu| &< 1.5 \times 10^{-19} \, e \, \text{cm} \, ,  
\end{align}
both at $95 \%$ C.L..

Analytical results for all the possible contributions to the magnetic and electric dipole moment can be found in~\cite{ultral}. However, we are going to concentrate on the lepton flavor conserving case, since too high values for the flavor violating couplings would be necessary to fully explain the anomalies. The analytical expressions, which we note that are not approximations, are given by
\begin{equation} \label{eq:anomalous_magnetic_moment_diag}
  \Delta a_{\alpha} = \frac{1}{16 \pi^{2}} \left[ 3 \, \left( \text{Re} \, S^{\alpha \alpha} \right)^2 - \left( \text{Im} \, S^{\alpha \alpha} \right)^2 \right] \, ,
\end{equation}
\begin{equation} \label{eq:EDM_diag}
  d_{\alpha} = - \frac{e}{8 \, \pi^2 \, m_\alpha} \left( \text{Re} \, S^{\alpha \alpha} \right) \, \left( \text{Im} \, S^{\alpha \alpha} \right) \, .
\end{equation}
Notice here that in the case of a pseudoscalar particle, which is the case in the majority of Goldstone bosons examples, the contribution to the g-2 is always negative and the muon anomaly cannot be explained.

\begin{figure}
  \begin{subfigure}{0.5\textwidth}
    \centering
    \includegraphics[scale=0.4]{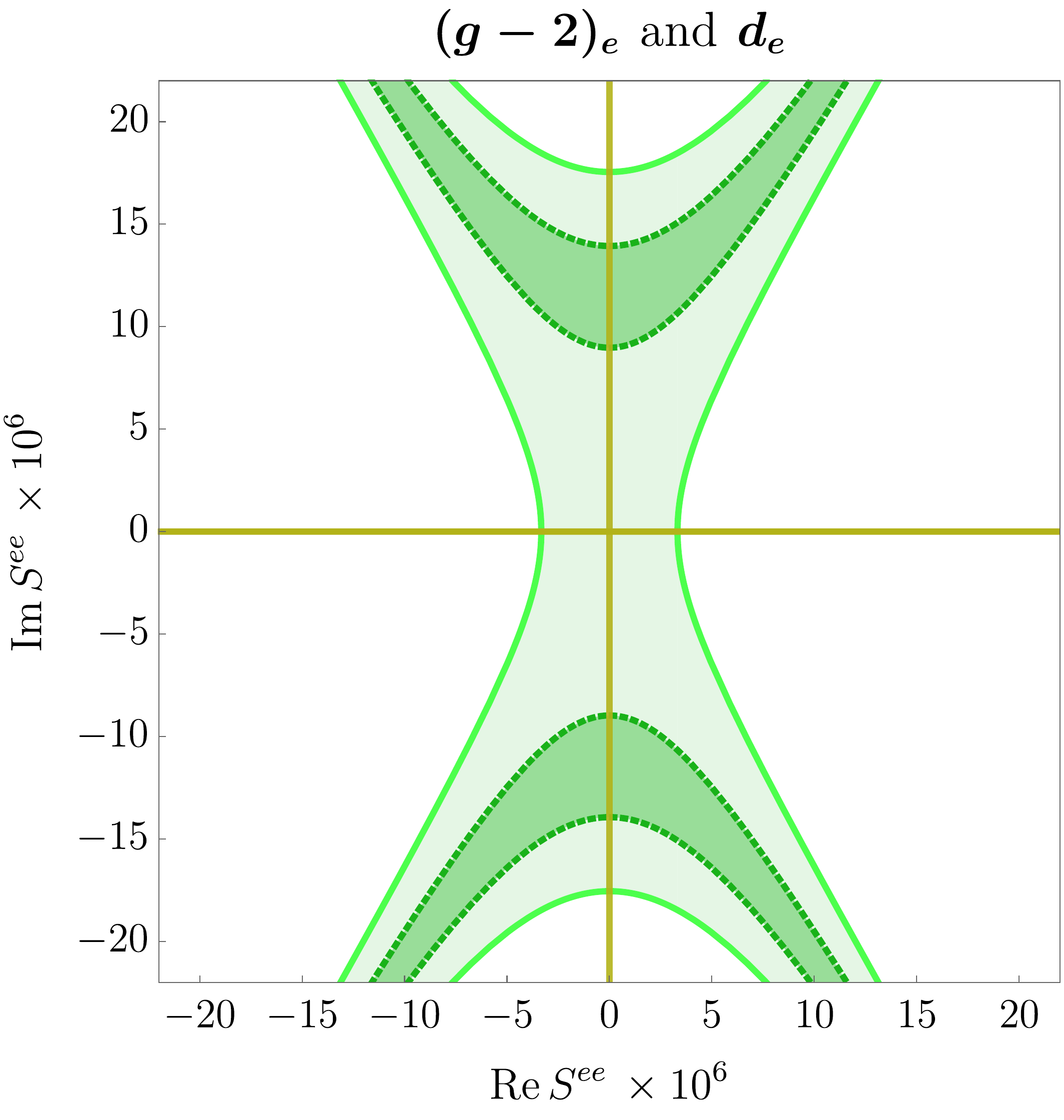}
  \end{subfigure}
  \begin{subfigure}{0.5\textwidth}
    \centering
    \includegraphics[scale=0.4]{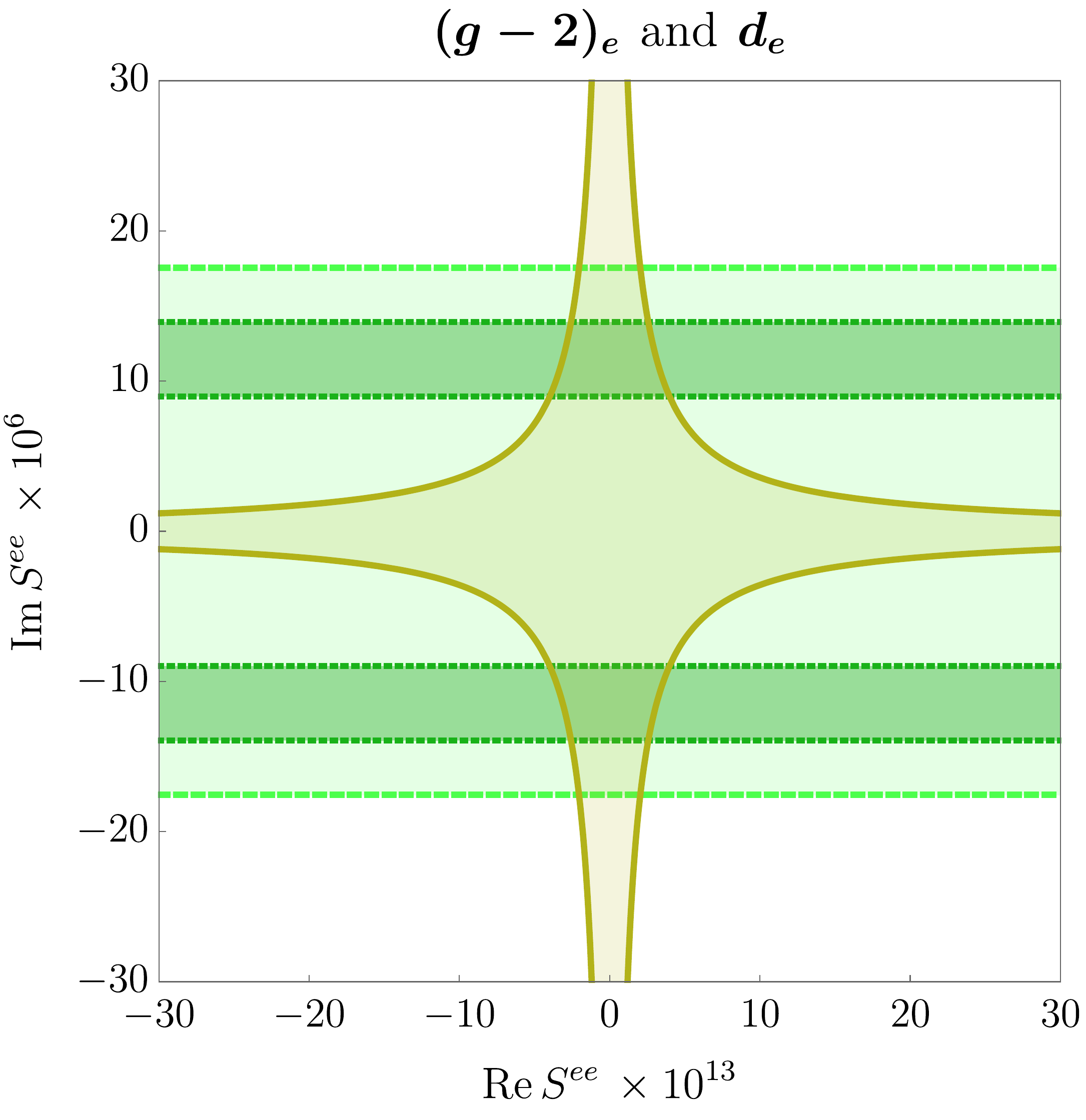}
  \end{subfigure}
  \caption{Preferred regions for the $\phi - e - e$ coupling due
    to the electron AMM and EDM. The deviation in the
    electron AMM is explained at the $3\,\sigma$ ($1\,\sigma$)
    level inside the light (dark) green area. The region delimited by the orange continuous lines is the
    region allowed by the current experimental limit of
    the electron EDM. The same plot is shown in the figure on the right with the abscissa axis zoomed.
    \label{fig:eMoments}}
\end{figure}

\begin{figure}
  \begin{subfigure}{0.5\textwidth}
    \centering
    \includegraphics[scale=0.4]{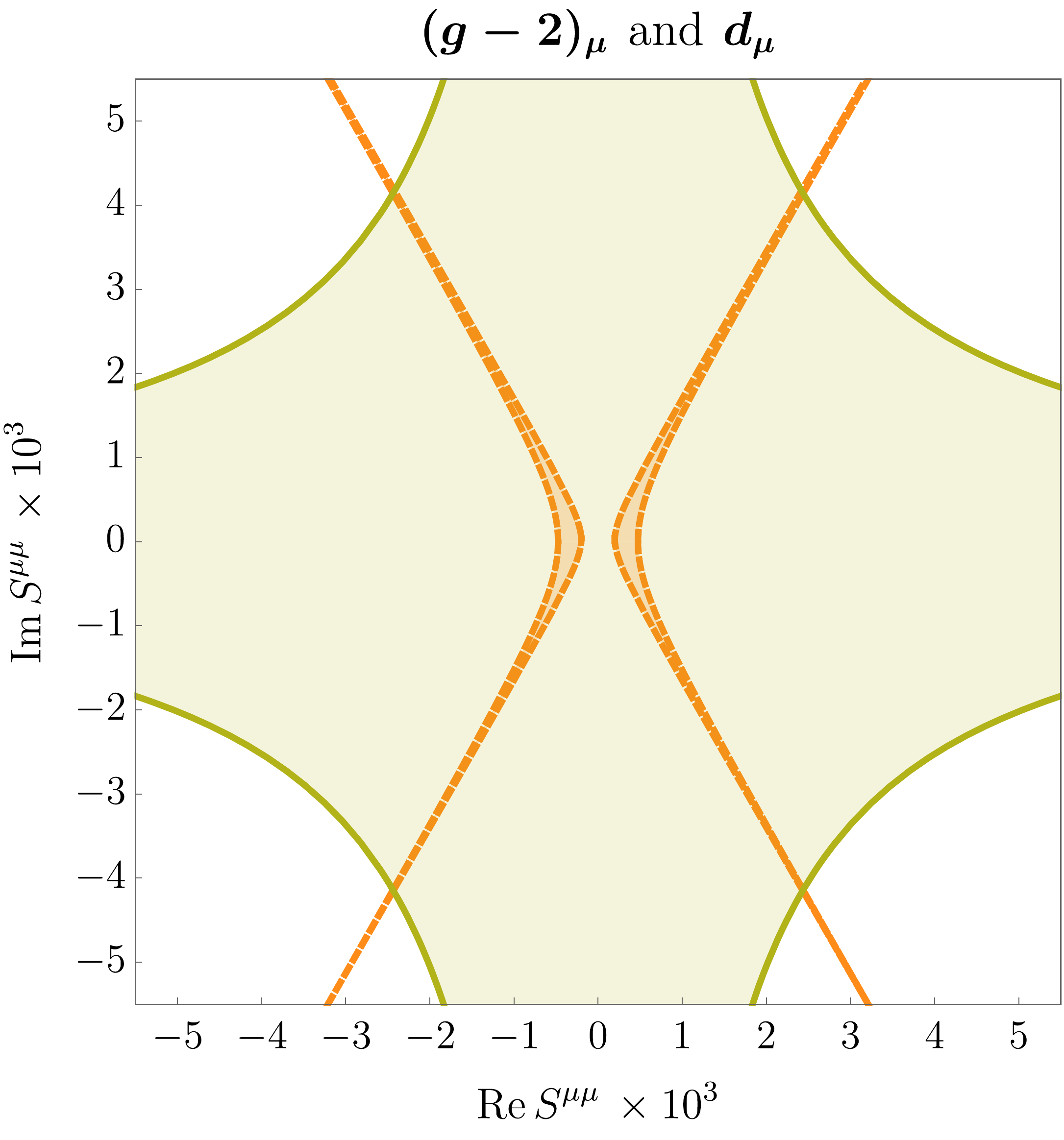}
  \end{subfigure}
  \begin{subfigure}{0.5\textwidth}
    \centering
    \includegraphics[scale=0.4]{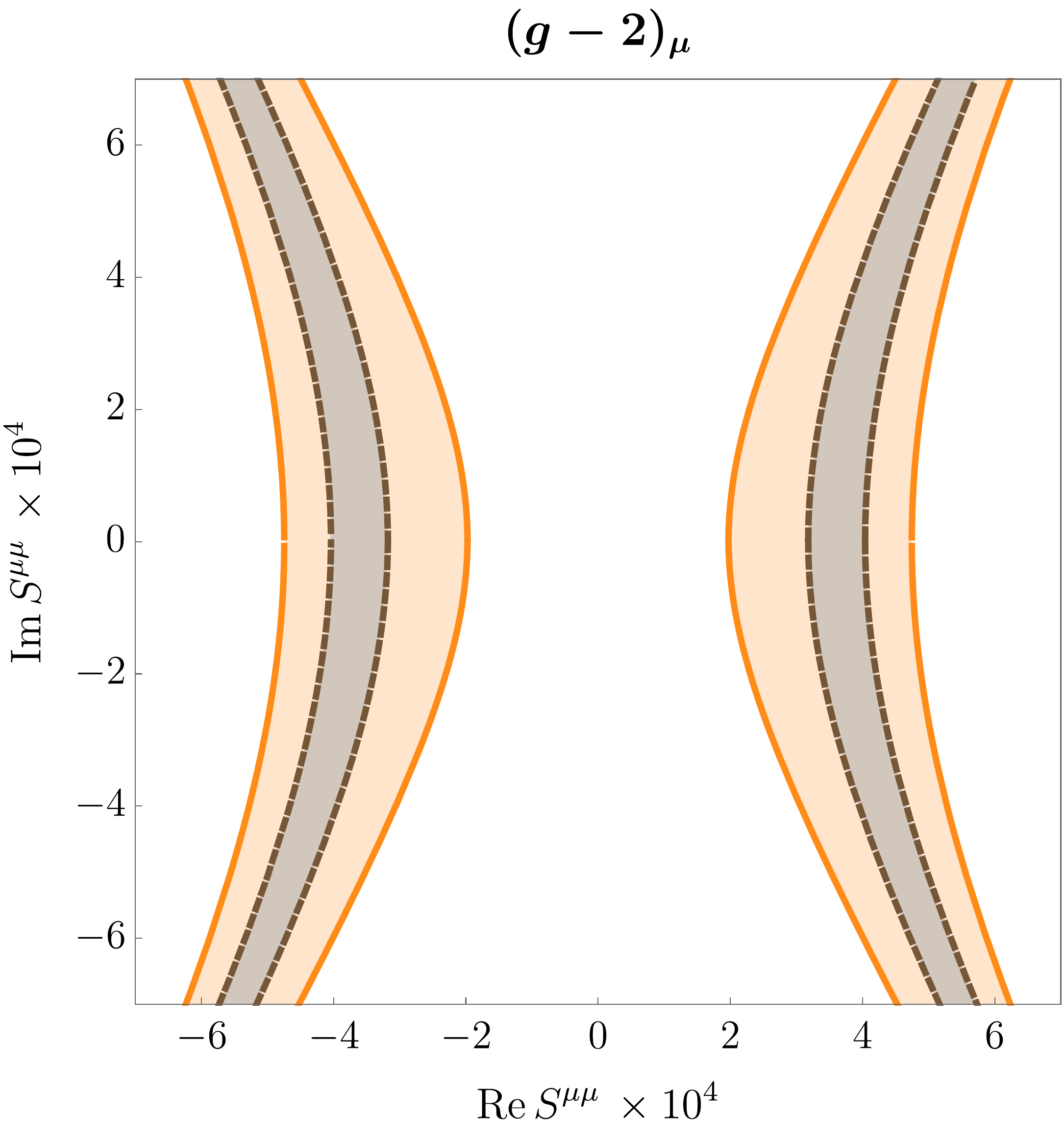}
  \end{subfigure}
  \caption{Preferred regions for the $\phi - \mu - \mu$ coupling due
    to the muon AMM and EDM. It can be seen in the figure on the left that the bound from the muon EDM
    (yellow continuous curves) does not restrict too much the coupling, being able to explain the AMM of the muon (orange dashed curves) in a wide range. On the right figure, only the
    muon AMM is represented and 
    the deviation in the
    muon AMM is explained at the $3\,\sigma$ ($1\,\sigma$)
    level inside the light (dark) area.
    \label{fig:muMoments}}
\end{figure}

Using the previous expressions, we show in Figure~\ref{fig:eMoments} favored regions for the electron diagonal coupling due to the electron AMM and EDM. Inside the light green region we can explain the g-2 anomaly at $3 \, \sigma$, while in the darker region it is explained at $1 \, \sigma$. On the left panel, it is clearly seen that the bound on the electron EDM, which is the orange region, strongly constrains the coupling, making it essentially purely real or essentially purely imaginary. However, given the low significance of the electron AMM deviation, we can find regions in the parameter space where the anomaly is explained. We can stay in the $3 \, \sigma$ region even with $S^{ee} = 0$, but if $\text{Re} \, S^{ee} \lesssim 10^{-13}$,  a value of $\text{Im} \, S^{ee} \sim 10^{-5}$ would place us in the $1\,\sigma$ region. On the other hand, one must introduce larger couplings in order to reconcile the theoretical prediction with the experimental measurement in the case of the $\left(g - 2\right)_\mu$, given thath the deviation is more significant here. This can be clearly seen in Figure~\ref{fig:muMoments}. Now, the bound from the muon EDM is not restrictive enough to appreciatly restrict the parameter space, as shown in the left panel. However, larger values of $S^{\mu \mu}$, of the order of $10^{-4}$, are needed to explain the current anomaly. However, in both cases, the required values of the couplings are in conflict with the bounds given in~\cite{ultral} for the diagonal couplings, 
\begin{equation}
  \IM \, S^{ee} < 2.1 \times 10^{-13} \quad , \quad \IM \, S^{\mu \mu} < 2.1 \times 10^{-10} \quad , \quad \RE \, S^{\beta \beta} \lesssim \left[ \IM \, S^{\beta \beta} \right]_{\max} \, .
\end{equation}
These limits come from astrophysical observations and are based on the assumptions that the scalar properties are the same in the astrophysical medium as they are in the vacuum, but some mechanisms have been recently proposed~\cite{bloch2021, derocco2021} under which these assumptions are invalid. Additional bounds not derived from astrophysical scenarios can also be found in~\cite{ultral},
\begin{equation}
  S^{ee} \lesssim 10^{-7} \quad , \quad S^{\mu \mu} \lesssim 10^{-5} \, .
\end{equation}
In particular, they are obtained using the results from the OSQAR experiment~\cite{osqar}. Therefore, a mechanism to suppress the processes from which the limits are derived would be necessary for the ultralight scalar to be able to fully explain the current g-2 anomalies.

\section{Summary and discussion}

A broad variety of SM extensions include ultralight scalars both in the form of exactly massless particles, as is the case of Goldstone bosons, and as states much lighter than any other massive particle in the model. 

In this work we have explored the impact of ultralight scalars adopting a model independent general approach, considering both scalar and pseudoscalar interactions. First, we have derived bounds on the lepton flavor violating couplings of the ultralight scalar with the charged leptons and we have explored some phenomenological aspects of this scenario. In particular we have seen that the observables discussed in the paper are complementary and also, that a full explanation to the $g-2$ anomalies can be possible if some mechanisms exist to suppress the processes from which the bounds on the diagonal couplings are obtained. However, this is not the case with pure pseudoscalars, since the contribution to the observable has the opposite sign as that of the muon anomaly.

Finally, since ultralight scalars can appear in most high- and low energy processes, their phenomenology is very rich. For instance, these scalars can also couple to quarks, opening many hadronic and semi-leptonic channels. Therefore, in our opinion, the ultralight scalars deserve further investigation due to the wide diversity of experimental possibilities that they include.

\section*{Acknowledgements}

This article is based on the talk given in the BSM-2021 at Zewail City. The original work~\cite{ultral} was done in collaboration with Avelino Vicente, whom I would like to thank for his help writting this manuscript. It was supported by the Spanish grants
FPA2017-85216-P (MINECO/AEI/FEDER, UE), FPA2017-90566-REDC (Red Consolider MultiDark) and by the FPI grant PRE2018-084599.

\bibliographystyle{unsrt}

\begin{thebibliography}{99}
\bibitem{cs18} L. Calibbi and G. Signorelli, Riv. Nuovo Cim. {\bf 41} no. 2, (2018) 71-174.
\bibitem{meg} {\bf MEG II} Collaboration, A. Baldini {\it et al.}, Eur. Phys. J. C. {\bf 78} no. 5, (2018) 380.
\bibitem{papa} A. Papa, EPJ Web Conf. {\bf 234}, (2020) 01011.
\bibitem{mu3e} {\bf Mu3e} Collaboration, N. Berger, Nucl. Phys. B Proc. Suppl. {\bf 248-250} (2014) 35-40.
\bibitem{g2muFermi} {\bf Muon g-2} Collaboration, B. Abi {\it et al.}, Phys. Rev. Lett. {\bf 126} (2021) 141801.
\bibitem{g2muBrook} {\bf Muon $\mathbf{g - 2}$} Collaboration, G. Benett {\it et al.}, Phys. Rev. D {\bf 73} (2006) 072003.
\bibitem{Bennett2008dy} {\bf Muon $\mathbf{g - 2}$} Collaboration, G. Benett {\it et al.}, Phys. Rev. D {\bf 80} (2009) 052008.
\bibitem{porod14} W. Porod, F. Staub and A. Vicente, Eur. Phys. J. C {\bf 74} no. 8, (2014) 2992.
\bibitem{ultral} P. Escribano and A. Vicente, JHEP {\bf 03} (2021) 240.
\bibitem{hirsch09} M. Hirsch, A. Vicente, J. Meyer and W. Porod, Phys. Rev. D {\bf 79} (2009) 055023.
\bibitem{Jodidio:1986mz} A. Jodidio {\it et al.}, Phys. Rev. D {\bf 34} (21986) 1967.
\bibitem{twist} {\bf TWIST} Collaboration, R. Bayes, Phys. Rev. D {\bf 91} (2015) 052020.
\bibitem{gouvea2013} A. de Gouvea and P. Vogel, Prog. Part. Nucl. Phys. {\bf 71} (2013) 75-92.
\bibitem{Tanabashi2018oca} {\bf Particle Data Group} Collaboration, M. Tanabashi {\it et al.}, Phys. Rev. D {\bf 98} no. 3, (2018) 030001.
\bibitem{Perez2019cdy} {\bf Belle-II} Collaboration, D. Rodríguez Pérez, 17th Conference on Flavor Physics and CP Violation. 6, 2019.
\bibitem{Andreev2018ayy} {\bf ACME} Collaboration, V. Andreev, Nature {\bf 562} (2018) 355-360.
\bibitem{bloch2021} I. M. Bloch, A. Caputo, R. Essig, D. Redigolo, M. Sholapurkar and T. Volansky, arXiv:2006.14521 [hep-ph].
\bibitem{derocco2021} W. DeRocco, P. W. Graham and S. Rajendran, arXiv:2006.15112 [hep-ph].
\bibitem{osqar} R. Ballon {\it et al.}, 10th Patras Workshop on Axions, WIMPs and WISPs, pp. 125-130. 2014.
\bibitem{Mori2016vwi} {\bf MEG} Collaboration, T. Mori, Nuovo Cim. C {\bf 39} no. 4, (2017) 325.
\bibitem{Bolton1986tv} R. Bolton {\it et al.}, Phys. Rev. Lett. {\bf 56} (1986) 2461-2464.
\bibitem{Goldman1987hy} J. Goldman {\it et al.}, Phys. Rev. D {\bf 36} (1987) 1543-1546.
\bibitem{Bolton1988af} R. Bolton {\it et al.}, Phys. Rev. D {\bf 38} (1988) 2077.

\end{thebibliography}

\end{document}